\input{aipcheck}

\documentclass[
    ,final            
  ]
  {aipproc}

\layoutstyle{8x11single}

\begin{document}

\title{Latest News from Double Chooz Reactor Neutrino Experiment}

\author{Masahiro Kuze for Double Chooz Collaboration}{
  address={Department of Physics, Tokyo Institute of Technology, Tokyo 152-8551 Japan}
}



\begin{abstract}
Double Chooz experiment will search for a disappearance of
the reactor neutrinos from Chooz reactor cores in Ardenne, France, in order to
detect the yet unknown neutrino oscillation angle $\theta_{13}$.
The far detector was completed in 2010 and data-taking has started in
spring 2011.  Status of data-taking is presented  and some
performance plots from physics data are shown in this paper
for the first time.  Also the prospect of experimental sensitivity
is presented, in light of recent indication from T2K for a non-zero
$\theta_{13}$ value.
\end{abstract}

\maketitle


\section{Introduction}
Unlike the quark sector, the nature of lepton mixing phenomenon
is still poorly revealed.  Especially, while two other angles are
large, the mixing angle $\theta_{13}$ seems to be small and only
an upper limit $\sin^2 2\theta_{13} < 0.15$ has been set~\cite{CHOOZ}.
Measurement of this angle is one the most important tasks of neutrino
physics in next few years, as the value of $\theta_{13}$ determines the
detectability of CP violation phase $\delta_{CP}$ in future experiments.

Oscillation probability for mixing angle $\theta$ and mass difference
$\Delta m^2 (\rm eV^2)$ is $\sin^2 2\theta \sin^2 \left(
1.27 \Delta m^2 L / E \right)$, where $L$ is distance (km) and $E$ is
neutrino energy (GeV).  For $\Delta m^2_{31} = 2.5 \times 10^{-3} \rm eV^2$,
this corresponds to first oscillation maximum
at a few 100 km for accelerator neutrino ($E \approx$  1~GeV), while it is
1$\sim$2~km for reactor neutrino of $E \approx$ 4~MeV.
(For the same energy, KamLAND explores the sector  $\Delta m^2_{21} = 
8 \times 10^{-5} \rm eV^2$ at much larger distances.)

Therefore, the principle of reactor $\theta_{13}$ experiment is to measure
the electron anti-neutrino flux precisely at a distance of 1$\sim$2 km from the source
and detect disappearance (the energy is not enough for appearance
of other leptons).
Since it is a pure $\theta_{13}$ measurement, it is complementary to
accelerator appearance experiment and both methods are necessary
to yield richer physics results.
To reduce the uncertainties related to source flux estimation, new-generation
reactor experiments (Daya Bay in China, Double Chooz in France and
RENO in Korea) have a concept of placing identical detectors at near
(before oscillation) and far locations and make a ratio of the
two measurements.  

All these experiments have a liquid scintillator (LS) target doped with Gd,
using the Inverse-$\beta$ reaction $\bar \nu_e  p \to e^+ n$.
Gd has a very large neutron absorption cross section.
The positron deposits its kinetic energy and annihilates in the target,
giving rise to a prompt signal.  The neutrino energy can be measured
from this signal.  The neutron drifts in the target and thermalizes,
finally absorbed by Gd which emits a few $\gamma$ rays with a total
energy of 8~MeV.  This signal occurs $\sim 30~\mu \rm s$ after the
prompt signal on average and is called the delayed signal.
Making "delayed coincidence" of these two signals is a key to
separate a neutrino signal from the large backgrounds from
environment and cosmic rays.

\section{Double Chooz experiment}
The details of the experiment are described elsewhere~\cite{Proposal}.
The detector has four layers of cylindrical vessels: the neutrino
target (Gd-doped LS), Gamma-catcher (LS without Gd) which detects
$\gamma$ rays from neutrino interaction that escaped from the target,
Buffer layer (non-scintillating mineral oil) to shield the active region
from external $\gamma$ rays mainly from photomultiplier (PMT) glass,
and  Inner Veto (LS) to veto cosmic rays.  Inner Detector (up to Buffer layer)
and Inner Veto are optically separated and viewed by 390 and 78 PMTs,
respectively.

In Dec. 2010, filling of liquids in the far detector was finished and the 
main far detector was completed.  From Jan. 2011, commissioning of
the detector (such as HV tuning) was done and physics data-taking
started from Apr. 2011.  In parallel, installation of Outer Veto (plastic
scintillator strips to track cosmic muons) is on-going and its lower
part has already been completed.  Glove Box for calibration (to deploy
radioactive and other sources) has just been installed.

For the near detector, civil construction of the tunnel and the lab hall
is ongoing, and will be finished in Apr. 2012.  The detector construction
will then follow, expected to be finished by the end of 2012.  After that,
the data-taking will be in "double" mode with reduced systematic
uncertainties.

Stable data-taking for physics started on 13 Apr. with on-site shifters
and remote shifts in three time-zones.  Data acquisition efficiency is
more than 85\% of the time.  It includes calibration runs with light injection
by embedded fibers.  Trigger rate is about 120~Hz.  So far, already 70
days of physics data have been accumulated.  Still detailed checks on
data are to be made, but preliminary plots from physics runs are shown
in the next section.

\section{Data from Physics Runs}
\begin{figure}
  \includegraphics[width=.475\columnwidth]{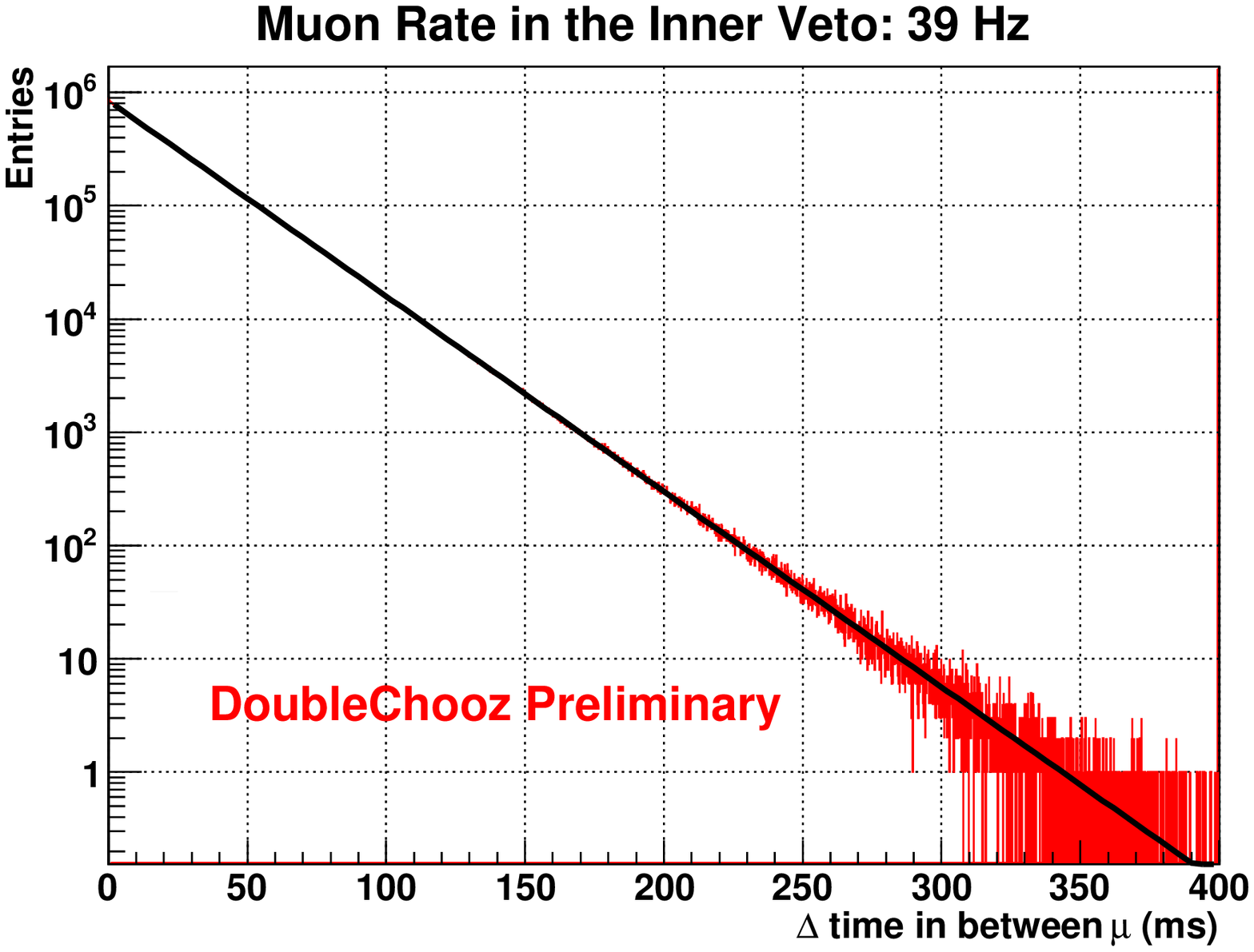}
\includegraphics[width=.5\columnwidth]{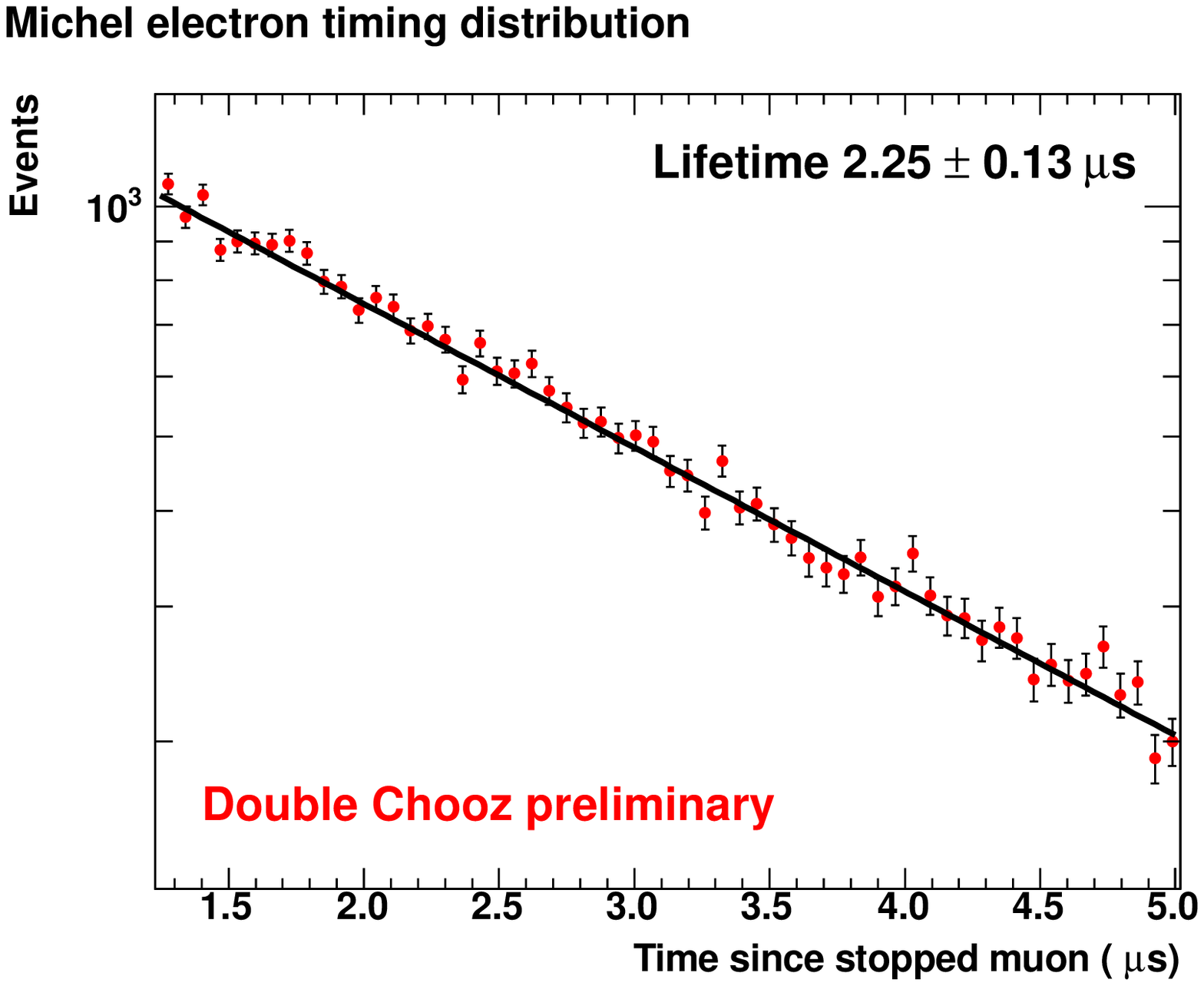}
   \label{muon}
 \caption{(left) Distribution of time between two muon events in Inner Veto.
 (right) Michel electron timing distribution.}
\end{figure}

Figure~\ref{muon}(left) shows the time distribution between muon events tagged by
Inner Veto.  About 40~Hz of muons are detected, while about 10~Hz of muons
are detected in Inner Detector.  Figure~\ref{muon}(right) shows the Michel electron
timing distribution from stopped muons.  The life time is consistent with
muon decay, demonstrating that delayed coincidence technique is working well.

\begin{figure}
 \includegraphics[width=.5\columnwidth]{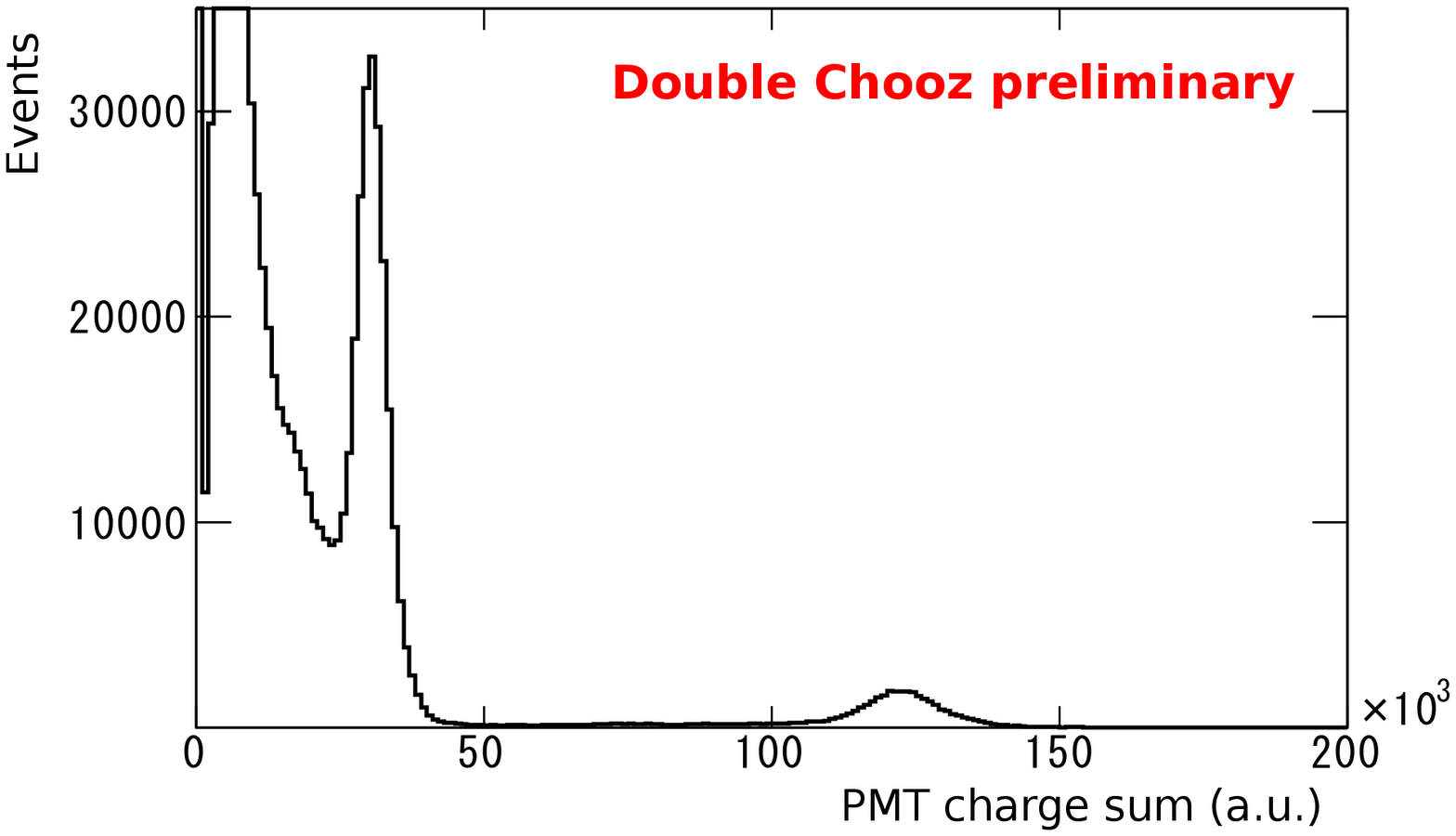}
 \includegraphics[width=.5\columnwidth]{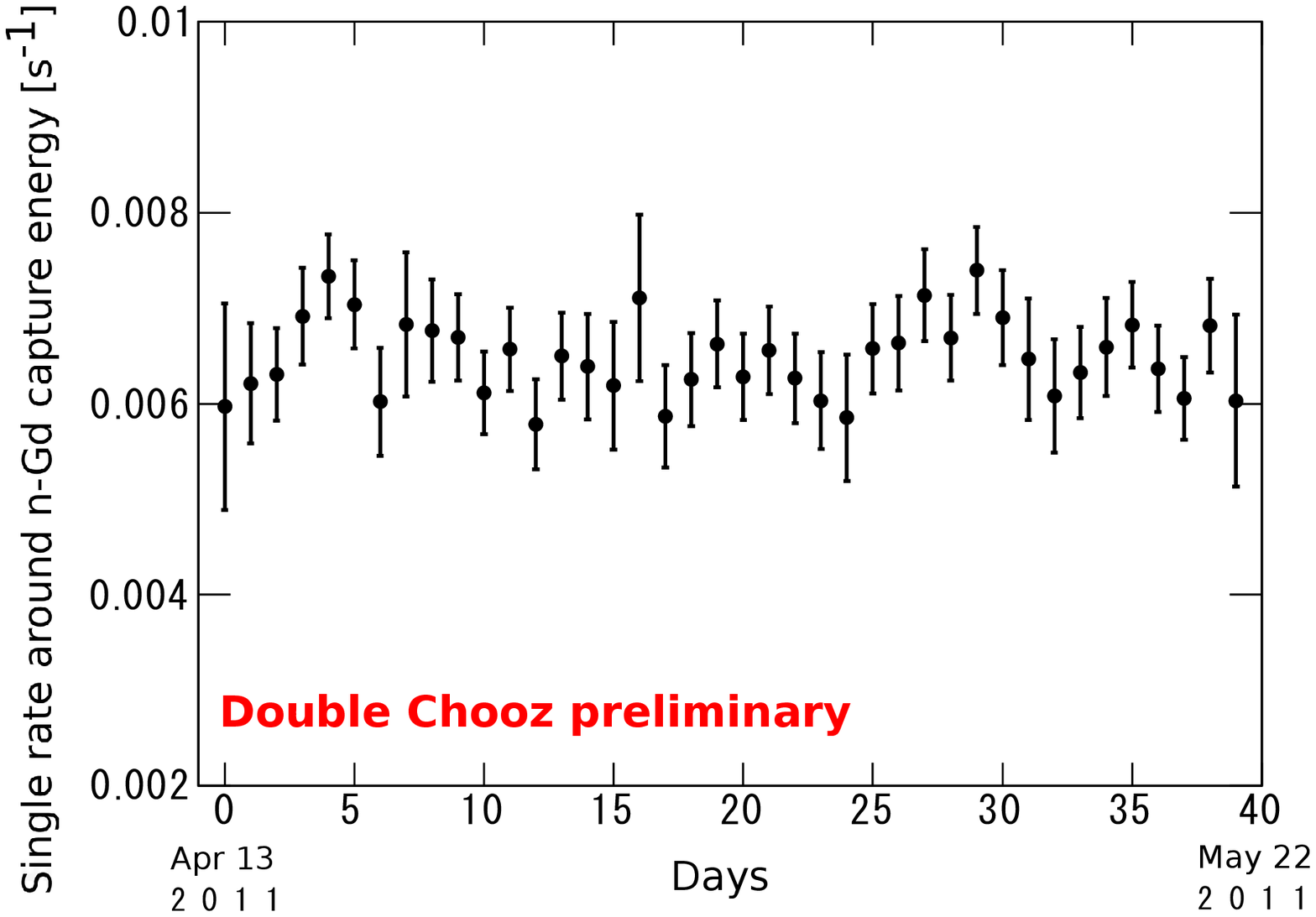}
   \label{neutron}
 \caption{(left) Delayed energy in muon-correlated time window.
 (right) Singles rate in delayed energy window vs. date.}
\end{figure}

Figure~\ref{neutron}(left) shows the delayed energy distribution in muon-correlated
time window.  Note that it is from raw data, without gain calibration nor energy
calibration/correction.  Clean peaks are seen for neutron capture by H (2.2~MeV)
and Gd (8~MeV).  Events in H capture includes interactions in $\gamma$-catcher
volume.  Figure~\ref{neutron}(right) shows the stability of singles rate in delayed energy
window (currently 6 to 12~MeV), after vetoing muon-correlated events.  
The rate is well below 0.01~Hz,
which is less than half of the proposal value (0.024~Hz, which was scaled from Chooz
experiment).  The singles rate in prompt energy window (currently 0.7 to 12~MeV)
is 10~Hz, as was expected in the proposal and much lower than previous Chooz
experiment (64.8~Hz) thanks to the new Buffer layer.

From these observed performances, the accidental coincidence rate is expected
to be small enough and a clean set of neutrino candidates is anticipated in the current
data.  Detailed studies on correlated backgrounds (fast neutrons and long-lived
isotopes induced by cosmic rays) will also be made.

\section{Prospects}
\begin{figure}
 \includegraphics[width=.4\columnwidth]{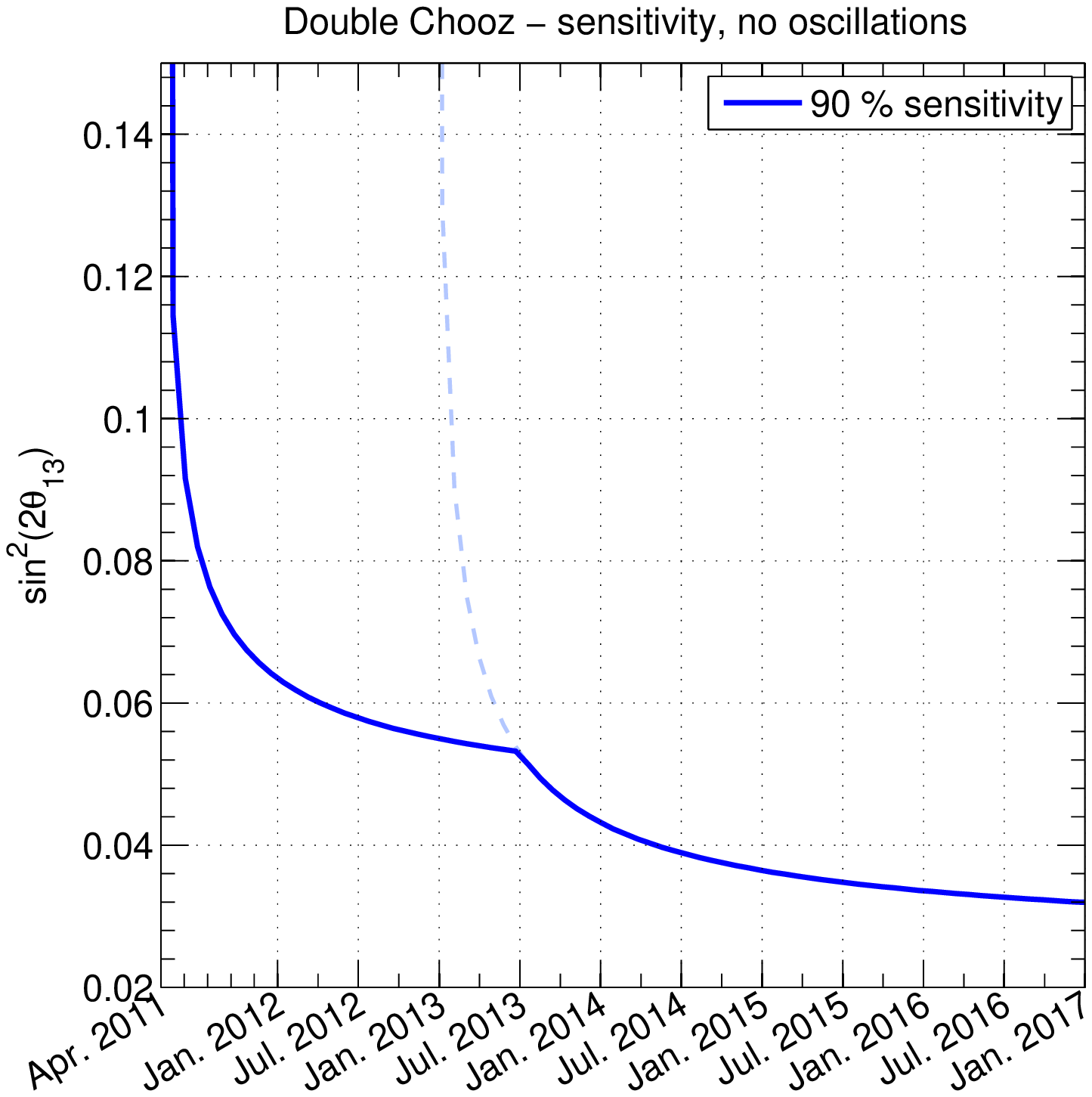}
  \includegraphics[width=.4\columnwidth]{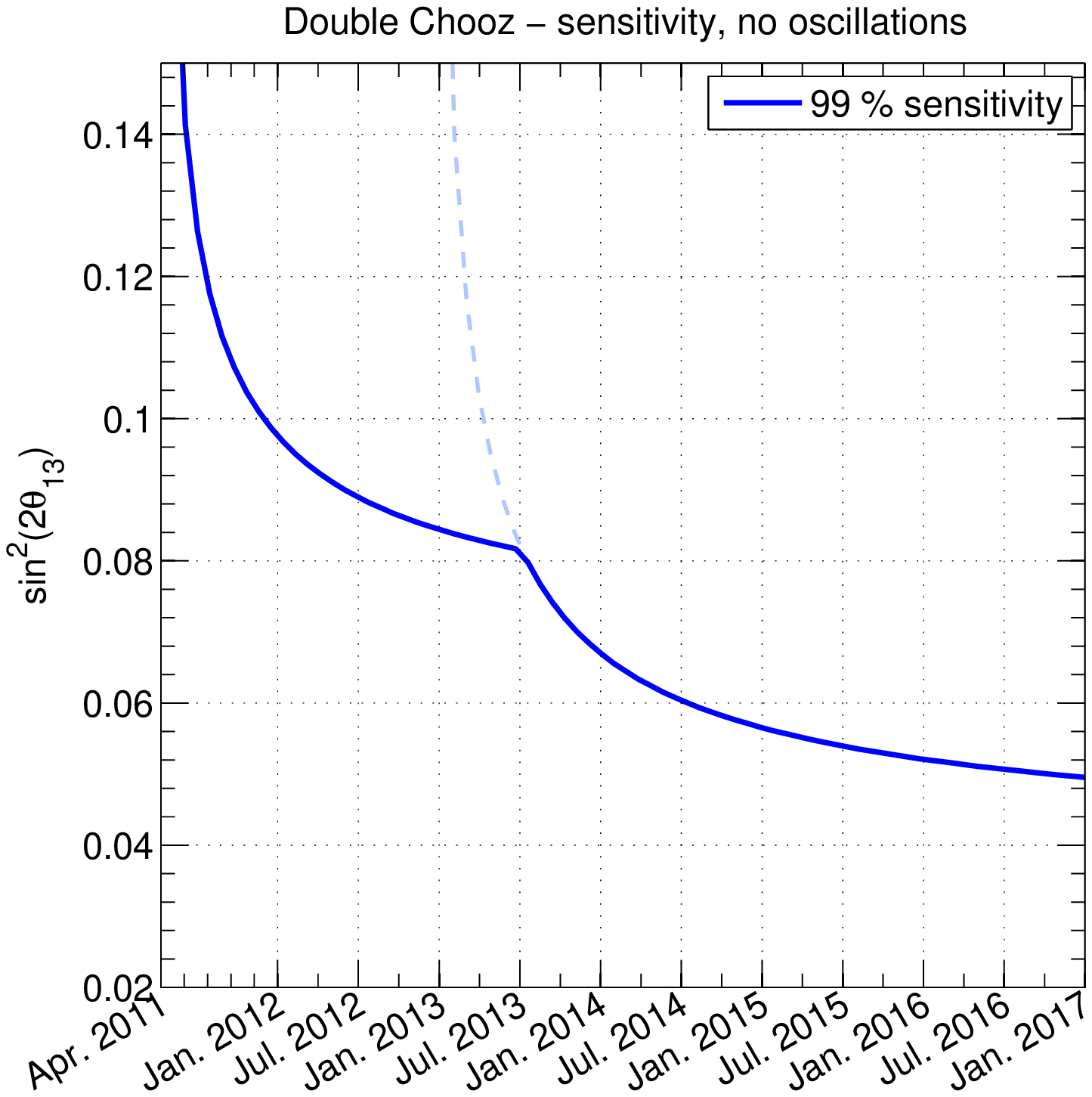}
    \label{sens}
 \caption{Sensitivity on $\sin^2 2\theta_{13}$ as a function of time at (left) 90\% and (right) 99\%
 Confidence Level.}
\end{figure}
Figure~\ref{sens} shows the revised sensitivity of Double Chooz as a function of time,
in which the start of near-detector data-taking is set to end of 2012.
Physics data-taking efficiency of 80\% and reactor duty-cycle of 80\% are assumed.
Both rate and spectral shape information of neutrino events are used in oscillation analysis.
For the far-detector only phase, normalization error of 2.5\% and bin-to-bin uncorrelated
error of 2\% are assumed.

Recently T2K collaboration observed an indication of electron appearance from
muon neutrino beam~\cite{T2K}.  Their result converts to confidence interval of 
$0.03(0.04) < \sin^2 2\theta_{13} < 0.28(0.34)$ at 90\% C.L. for normal (inverted) hierarchy.
The best-fit point is $\sin^2 2\theta_{13} = 0.11(0.14)$.  It can be seen that
T2K's best-fit values can be addressed at 99\% C.L. with Double Chooz 2011 data,
and almost all 90~\% C.L. interval can be covered by Double Chooz at 90\% C.L.
in full data set.

\section{Summary}
Double Chooz experiment started to take physics data.
The challenging "four-layer vessel" detector concept, developed by Double Chooz
colleagues around 2002, has proved to work after the construction period of 2008 - 
2010.  Still detailed checks and calibrations are to be made, but the performance
from raw data already looks promising towards observing a nice neutrino signal.
Given a hint from T2K for non-zero $\theta_{13}$, the interplay between reactor
and accelerator measurements will bring exciting time in neutrino physics.

\IfFileExists{\jobname.bbl}{}
 {\typeout{}
  \typeout{******************************************}
  \typeout{** Please run "bibtex \jobname" to optain}
  \typeout{** the bibliography and then re-run LaTeX}
  \typeout{** twice to fix the references!}
  \typeout{******************************************}
  \typeout{}
 }

\end{document}